\documentclass[preprint]{elsarticle}
\usepackage[utf8]{inputenc}
\usepackage[english]{babel}
\usepackage{amssymb}
\usepackage{amsmath}
\usepackage{multirow}
\usepackage{textcomp}
\usepackage{gensymb}

\usepackage{lineno}
\usepackage{setspace}
\begin{document}
\begin{frontmatter}

\title{Yield stress of aerated cement paste}

\author[navier,uio]{Blandine Feneuil\corref{cor1}}
\ead{bffeneui@math.uio.no}

\author[navier]{Nicolas Roussel}

\author[navier]{Olivier Pitois}

\cortext[cor1]{Corresponding author}

\address[navier]{Laboratoire Navier, UMR 8205, École des Ponts ParisTech, IFSTTAR, CNRS, UPE, Champs-sur-Marne, France}

\address[uio]{Current position : Department of Mathematics, University of Oslo, Oslo, Norway}

\begin{abstract}

Yield stress of aerated cement paste is studied. Samples are prepared by mixing aqueous foam with cement paste, which allows controlling bubble size, gas volume fraction and yield stress of the cement paste. Two distinct behaviors are observed depending on the surfactant used to prepare the precursor aqueous foam: (i) For a surfactant with low adsorption ability with respect to cement grains, bubbles tend to decrease the yield stress of the paste with magnitude governed by the Bingham capillary number, which accounts for bubble deformability. (ii) For a surfactant with high adsorption ability, bubbles increase significantly the yield stress. This behavior is shown to result from the surfactant-induced hydrophobization of the cement grains, which adsorb at the surface of the bubbles and tend to rigidify them. Within this regime, the effect of air incorporation is comparable to the effect of added solid particles.

\end{abstract}

\begin{keyword}

rheology (A) \sep cement paste (D) \sep bubbles
\end{keyword}

\end{frontmatter}

\newpage

\section{Introduction}

Yield stress of a cement paste, mortar or concrete is a crucial property. For instance, sufficient yield stress can stop bleeding~\cite{2017_Massoussi}. When material is poured in a formwork, yield stress can have negative effects if it prevents proper filling of the mold~\cite{2017_Massoussi} but it helps to reduce the lateral formwork pressure~\cite{2006_Ovarlez}. In the case of mortar sprayed on a vertical support, yield stress dictates the maximum thickness of the layer~\cite{2017_Gorlier_4}. Among cementitious materials, aerated materials raise growing interest. Entrainment of air bubbles  into concrete, up to 10\% by volume, is known to improve its durability in environments exposed to freeze-thaw cycles~\cite{2016_Aitcin_air}. At higher air content, aerated construction materials are promising for various industrial applications thanks to their low densities, low raw material needs, and improved thermal and acoustic properties.

The goal of this paper is to understand the effect of air bubbles on the yield stress of cement pastes containing air below 40\% by volume. These materials will be referred to as "aerated cement paste" or "bubbles suspensions in cement paste" as the word "foams" usually refers to materials with higher air volume content, above 64\%~\cite{2013_Cantat}, where bubbles are densely packed and deformed by their neighbors. In foams, bubble interactions induce specific rheological properties \cite{2013_Cantat}, that are not discussed in this paper. 

Various observations have been reported in the literature concerning the rheological properties of aerated cement pastes or mortars. \textit{A\"itcin}~\cite{2016_Aitcin_air} notes that entrained bubbles improve workability of concrete whereas \textit{Rixom and Mailvaganam}~\cite{1999_Rixom} reports a large increase of viscosity with the amount of entrained air. These opposite results may arise from the different experimental protocols and the different paste formulations. Actually, quantifying the effect of air inclusions on cement paste rheology, independently from the paste formulation, requires to determine normalized rheological quantities, e.g. yield stress  $\tau_y$ of the aerated paste  divided by the yield stress $\tau_{ref}$ of the suspending paste. Micromechanical analysis shows that the normalized yield stress is a function of air volume content $\Phi$ and of the deformation of the bubbles when the fluid is yielding~\cite{2008_Chateau,2013_Nguyen}. Deformability of the bubbles is characterized by the Bingham capillary number $Ca_y$, which compares the suspending fluid yield stress $\tau_{ref}$ and the bubble capillary pressure $2\gamma/R$:

\begin{equation}
Ca_y=\dfrac{\tau_{ref}}{2 \gamma/R}
\label{equation_Ca_y}
\end{equation}
where $\gamma$ is the air-liquid surface tension and $R$ is the bubble radius.
Such a behavior has been confirmed experimentally when $Ca_y\rightarrow \infty$ and $Ca_y\rightarrow 0$~\cite{2013_Kogan,2015_Ducloue} for model yield stress fluids, i.e. aerated concentrated emulsions.

The case of aerated cement paste is much more complex due to the potentially strong effect of added surfactants on the suspending fluid yield stress $\tau_{ref}$~\cite{2017_Feneuil}. In order to tackle this issue we prepare aerated cement paste by incorporating aqueous foam into cement paste within conditions where both water and surfactant amounts are controlled. In the first part of the study, the effects of both water and surfactants on the suspending fluid yield stress are quantified. The dimensionless yield stress $\tau_y/\tau_{ref}$ of the prepared aerated cement pastes is then assessed as a function of bubble size and air volume content for two different surfactants.

\section{Materials and methods}
\subsection{Materials} 
We use a CEM I cement from Lafarge, Saint Vigor factory. Specific surface area is 0.359~m$^2$/g and chemical composition is given in Table~\ref{table_chimie_ciment_c3}.

\begin{table}[!ht]
\begin{center}
\begin{tabular}{|c c c c c c c c c|}
\hline
C$_3$S & C$_2$S & C$_3$A & C$_4$AF & CaO/SiO$_2$ & MgO & Na$_2$O +0.658 K$_2$O & SO$_3$ & Gypsum  \\ \hline  
62\% & 16\% &  2.1\% & 15\% & 3 & 1.1\% & 0.34 \%  & 2.6\% & 2.4\% \\ \hline 
\end{tabular}
\caption{Chemical composition of CEM I cement from Lafarge, Saint Vigor}
\label{table_chimie_ciment_c3}
\end{center}
\end{table}

Two surfactants will be used: TTAB (tetradecyltrimethylammonium bromide) is a cationic surfactant provided by Sigma-Aldrich, its molar mass is 336 g/mol. Bio-Terge$^{\text{\textregistered}}$ is an anionic surfactant provided by Stepan, its molar mass is 315 g/mol.

Interaction of both surfactants with the same cement as used in this paper have been previously studied~\cite{2017_Feneuil}. Adsorption isotherm of surfactants on cement grains have been measured and surfactant Critical Micelle Concentration (CMC) have been measured in DI water and in cement synthetic cement pore solution. For TTAB, CMC is 1.5 g/L in water and 0.5 g/L in cement pore solution. Regarding Bio-Terge, its CMC is 2 g/L in water and 0.5 g/L in cement pore solution. The main results of this paper will be recalled in the discussion.

\subsection{Method}

\subsubsection{Preparation of aerated and reference cement pastes}
\label{part_protocol_c3}

As the yield stress of cementitious materials depends on their shear history, the same time schedule has been followed for all samples. First, distilled water and cement at water-to-cement mass ratio $W/C_i$ are mechanically mixed for 3 min. Then, the paste is left at rest for 20 minutes to allow for the formation of sulfo-aluminates \cite{2015_Bessaies}. After 20 minutes, foam is added to the cement paste and carefully incorporated with a spoon. Hand mixing allows us to check visually that the bubbles size is kept constant during incorporation of the foam. Mixing duration is 4 minutes % check
, and it is carried out always by the same operator in order to be similar in all experiments. 30 minutes after initial cement paste mixing, the rheometer cup is filled and rheometry starts.

Unfoamed samples, the so-called "reference cement pastes", were prepared for comparison purpose with foamed samples. In these cases, the above-described procedure was followed except that pure surfactant solution was added (28 min after initial cement paste mixing) instead of foam.

The weight of water contained into the foam added to the cement paste is measured in order to calculate the new effective water-to-cement ratio, $W/C_f$, which characterizes the suspending cement paste in the aerated sample. Final air content $\Phi$ is assessed by weighting the aerated paste in the rheometer cup. Indeed, the density $\rho$ of the sample is related to the air content by $\rho(\Phi)= \rho_0(1-\Phi)$, where $\rho_0=(W/C_f+1)/(W/C_f/\rho_w + 1/\rho_c)$ is the density of the suspending cement paste, with $\rho_w$ (resp. $\rho_c$) the density of water (resp. of cement). The maximal relative error on the measured air fraction is 4\%.

Note that in the chosen procedure, foam or surfactant solution is added more than 20 minutes after first mixing of cement and water, so surfactant is not expected to interfere in the formation of first hydration products~\cite{2015_Bessaies}.

\subsubsection{Foam production}
Aqueous foam is generated according to the method described in \cite{2019_Feneuil}. In brief, bubbles are created with a small T junction with two inputs: surfactant solution and nitrogen. Within such conditions, foams are made of bubbles having the same diameter, which is set by the flow rates of both nitrogen and surfactant solution. The foam is collected (bubble by bubble) in a column and it is wetted from the top to prevent it from drying and breaking. Once foam is ready to use, it is pushed from the column at constant flow rate.

\subsubsection{Samples}
We study seven sets of samples. For each set, both initial water-to-cement ratio (i.e. before addition of foam) and foam characteristics (surfactant and bubble size) are kept constant.

\begin{table}[!ht]
\begin{center}
\begin{tabular}{|c|c|c|c|c|c|}
\hline
 & Surfactant & Surfactant  & Bubble  & $W/C_i$  & Reference \\
 & &  concentration & diameter & & set \\\hline
Set T1 & TTAB & 4.5 g/L & 475 +/- 10 $\mu$m & 0.382 & R\_T1\\ \hline
Set T2 & TTAB & 4.5 g/L & 270 $\mu$m & 0.335 & R\_T2\\ \hline
Set T3 & TTAB & 4.5 g/L & 475 +/- 10 $\mu$m & 0.325 & R\_T3\\ \hline
Set T4 & TTAB & 4.5 g/L & 475 +/- 10 $\mu$m & 0.305 & R\_T4\\ \hline
Set T5 & TTAB & 4.5 g/L & 780 +/- 20 $\mu$m & 0.335 & R\_T2\\ \hline
Set B1 & Bio-Terge & 30 g/L & 440 +/- 5 $\mu$m & 0.329 & R\_B1\\ \hline
Set B2 & Bio-Terge & 8 g/L & 590 +/- 10 $\mu$m & 0.335 & R\_B2\\ \hline
 - & - & - & - & 0.305 to 0.405 & R\_0\\ \hline
\end{tabular}
\caption{Characteristics of each set of aerated cement paste samples. The corresponding reference paste for each set contains the same surfactant at the same concentration, and the same initial water-to-cement ratio.Note that sets T2 and  T5 share the same reference R\_T2.}
\label{table_samples}
\end{center}
\end{table}

Sets of reference samples are prepared to match with each set of bubble suspensions, i.e. they are prepared with the same surfactant solution and from the same initial cement paste, but they contain no bubble. In addition, reference set R\_0 refers to yield stress measurement on surfactant-free paste.

\subsubsection{Yield stress measurement}
Yield stress measurements are carried out with a vane-in-cup geometry in Ultra+ Kinexus Rheometer from Malvern. The six-blade vane tool is 5~cm high and 25~mm wide. We use a striated cup to avoid wall slip: diameter is 37 mm and height 62.5 mm. The cup is filled with aerated or reference cement paste with a spoon, then the Vane tool is slowly inserted into the paste. Pre-shear is prevented before measurement in order to avoid material flow and buoyancy-induced bubble migration~\cite{2006_Ovarlez,2008_Mahaut_a,2010_Goyon}.

Yield stress is measured with a start of flow sequence at a constant shear rate $\dot{\epsilon}=0.01 s^{-1}$ for 10 minutes~\cite{2008_Mahaut_b}. For all samples, stress increases up to the yield stress and then decreases. The yield stress is reached for shear strain values between 40 and 45\%. This value is low enough to enable correct measurement of yield stress with a Vane tool~\cite{2011_Ovarlez}. 

\begin{figure}[!ht]
\begin{center}
\includegraphics[width=6cm]{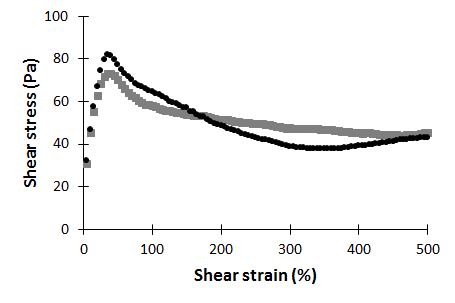}
\caption{Start of flow curves for yield stress measurement of an aerated cement paste (grey) and the corresponding reference paste (black). Examples correspond to set T3 with air content 31\% and the corresponding reference sample from set R\_T3.}
\label{curves_StartOfFlow}
\end{center}
\end{figure}

\section{Results}
 
\subsection{Reference yield stress}
\begin{figure}[!ht]
\begin{center}
\includegraphics[width=7cm]{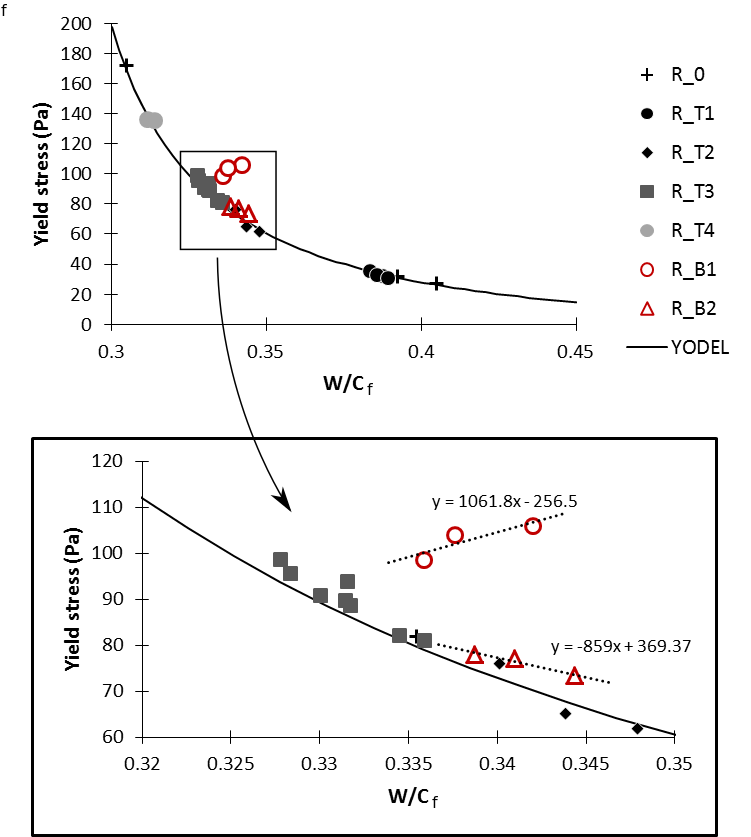}
\caption{Yield stress of reference cement pastes. Yield stresses for pastes with no surfactant (R\_0) or TTAB surfactant (R\_T1 to R\_T4) are fitted with equation \ref{equation_Yodel_c3} with $\Phi_{perc}=0.32$, $\Phi_{max}=0.545$ and $m_1=65~Pa$. In the case of Bio-Terge surfactant, linear regressions for each set of data are plotted with dotted lines.}
\label{curves_results_ref}
\end{center}
\end{figure}

Yield stresses for reference samples are plotted in Fig.~\ref{curves_results_ref}. As expected, when no surfactant is added, the yield stress decreases when the final water-to-cement ratio increases. For each reference data set containing TTAB surfactant, the yield stress decreases when the amount of added TTAB solution increases. The same trend is observed with Bio-Terge solution at the smallest concentration (R\_B2), even if the decrease is lower than for TTAB. On the contrary, when 30~g/L Bio-Terge surfactant solution is used (R\_B1), the yield stress increases with the amount of added solution.
    
\newpage

\subsection{Aerated cement paste}

\begin{figure}[!ht]
\begin{center}
\includegraphics[width=5cm]{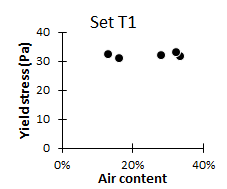}
\includegraphics[width=5cm]{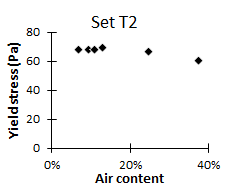}
\includegraphics[width=5cm]{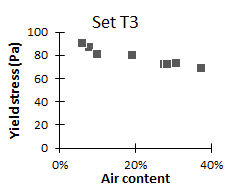}
\includegraphics[width=5cm]{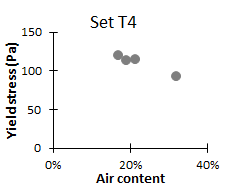}
\includegraphics[width=5cm]{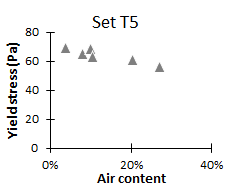}
\caption{Yield stresses of aerated cement pastes made with TTAB surfactant. Bubble diameter is 475~$\mu$m for T1,T3 and T4, 270~$\mu$mm for T2 and 780~$\mu$m for T5. Initial water-to-cement ratio is 0.382 for T1, 0.335 for T2 and T5, 0.325 for T3 and 0.305 for T4.} %Keep or remove data set descriptions in the caption ????
\label{curves_results_TTAB}
\end{center}
\end{figure}

\begin{figure}[!ht]
\begin{center}
\includegraphics[width=5cm]{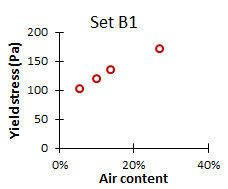}
\includegraphics[width=5cm]{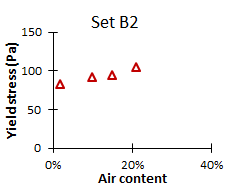}
\caption{Yield stresses of aerated cement pastes made with Bio-Terge surfactant. For B1, bubble diameter is 440~$\mu$m, initial water-to-cement ratio is 0.329, and surfactant concentration in foam is 30 g/L. For B2, bubble diameter is 590~$\mu$m, initial water-to-cement ratio, 0.335, and surfactant concentration in foam, 8 g/L.}
\label{curves_results_BioTerge}
\end{center}
\end{figure}

Yield stresses measured for each data set at different air contents are given in Figs.~\ref{curves_results_TTAB} and \ref{curves_results_BioTerge}. Evolution of yield stress with increasing air content is not the same for all data sets: it is constant in set T1, it decreases in sets T2 to T5 and increases for B1 and B2.

\section{Discussion}

\subsection{Reference yield stress with surfactant}

Ionic surfactants, mainly anionic, can strongly modify yield stress when they are added to a cement paste~\cite{2017_Feneuil}. At low concentration, adsorbed surfactant monolayer turns cement grains to hydrophobic grains which induces attractive interactions. At macroscopic scale, this results in the increase of the yield stress. At high surfactant concentration, surfactant agglomeration on the surface of the cement grains leads to steric repulsion between them and a drop of the paste yield stress. 

\begin{figure}[!ht]
\begin{center}
\includegraphics[width=6cm]{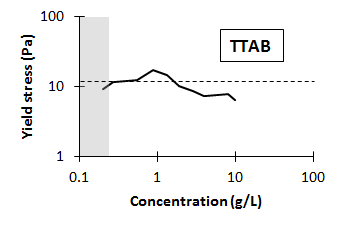}
\includegraphics[width=6cm]{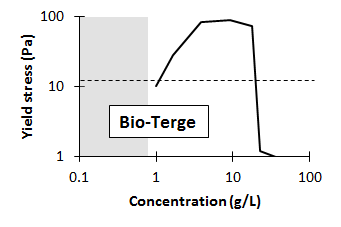}
\caption{Effect of surfactant concentration on the yield stress of cement pastes at $W/C_f=$0.5 prepared following the protocol described in~\cite{2017_Feneuil}. Repetition of a reference test in~\cite{2017_Feneuil} showed that the error is about 5\% of the yield stress value. Doted line shows yield stress of the paste with no surfactant (12~Pa). Grey areas show the surfactant concentrations used in the present study. }
\label{curves_concentration}
\end{center}
\end{figure} 

As surfactant concentration has a significant effect on the yield stress, let us calculate the final  concentrations $C_{TA,f}$ after mixing foam and cement paste. They depend on the concentration in the foaming solution $C_{TA,i}$, on the initial water-to-cement ratio of the precursor cement paste $W/C_i$ and the final water-to-cement ratio $W/C_f$:

\begin{equation}
C_{TA,f}=C_{TA,i}  \dfrac{W/C_f-W/C_i}{W/C_f}
\label{equation_concentration_TA}
\end{equation}

Calculated values range from 0.03 to 0.2~g/L for samples containing TTAB and between 0.08 and 0.8~g/L for Bio-Terge. In Fig.~\ref{curves_concentration}, yield stress is given as a function of surfactant concentration for cement pastes at $W/C_f=$ 0.5 prepared as in~\cite{2017_Feneuil}, with the same cement as this work, and containing TTAB or Bio-Terge. The figure shows that for both surfactants, the estimated concentrations are well below the concentrations at which maximum yield stresses are expected. Surfactants are therefore in the low concentration regime, where yield stress can be enhanced by hydrophobic interactions, but due to the very low concentrations used, we can expect this effect to be insignificant. Further analysis is made below for each surfactant.

\subsubsection{Reference yield stress with TTAB}
In Fig.~\ref{curves_results_ref}, all yield stresses of cement pastes containing no surfactant or TTAB  follow the same curve. First of all, this shows that TTAB, at the low concentrations we use here, has no effect on yield stress. In addition, yield stress depends only on the total water content in the cement paste, including both initial mixing water and foaming solution added later.

We choose to fit TTAB reference curve with equation \ref{equation_Yodel_c3}~\cite{2006_Flatt}:

\begin{equation}
\tau_{ref}=m_1 \dfrac{\Phi_p^2 (\Phi_p-\Phi_{perc})}{\Phi_{max} (\Phi_{max}-\Phi_p) }
\label{equation_Yodel_c3}
\end{equation}

In this equation, $m_1$ accounts for the intensity of the interactions between cement grains, whose major components are Van der Waals, electrostatic and steric forces~\cite{2006_Flatt}. $\Phi_p$ is the solid volume fraction  and is related to the water-to-cement ratio and to the densities of water $\rho_w$ and cement $\rho_c$:

\begin{equation}
\Phi_p=1/(1+\rho_c/\rho_w  W/C_f)
\label{equation_PhiP_WC}
\end{equation}
$\Phi_{perc}$ and $\Phi_{max}$  are respectively is the percolation threshold and the maximal solid fraction of the cement grains.

This curve is shown in Fig. \ref{curves_results_ref} with parameters $\Phi_{perc}=0.32$, $\Phi_{max}=0.545$ and $m_1=65~Pa$.

\subsubsection{Reference yield stress with Bio-Terge}

When Bio-Terge solution is added to cement paste, yield stress is higher than for samples containing TTAB or for surfactant-free pastes at the same water-to-cement ratio. In fact, addition of Bio-Terge solution into a cement paste has two consequences: solid fraction $\Phi_p$ decreases and the intensity of of the attractive forces between cement grains increases.  As a result, reference yield stress $\tau_{ref}$ cannot be fitted by equation \ref{equation_Yodel_c3}. Therefore, we choose to describe the corresponding data with linear regressions: $\tau_{ref}=1062~W/C-257$ for set R\_B1 and $\tau_{ref}=369-859~W/C$ for set R\_B2.

\subsection{Dimensionless yield stress}

\begin{figure}[!ht]
\begin{center}
\includegraphics[width=8cm]{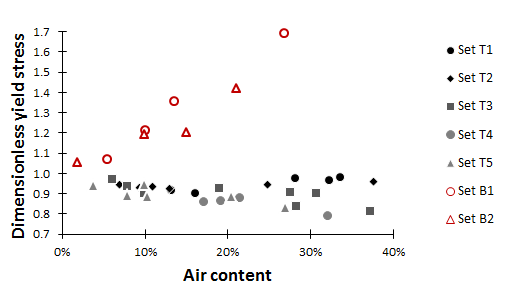}
\caption{Dimensionless yield stress for all samples (see Table \ref{table_samples} for characteristics of the samples).}
\label{curves_dimensionless_YS}
\end{center}
\end{figure}

Results for the normalized yield stress, i.e. $\tau_y/\tau_{y,ref}$, are shown in Fig. \ref{curves_dimensionless_YS}.

First of all, data corresponding to the different samples presented in Figs. \ref{curves_results_TTAB} and \ref{curves_results_BioTerge} collapse on two consistent sets characterized by the surfactant used. For Bio-Terge samples, the dimensionless yield stress increases with air content. On the other hand, with TTAB surfactant, all the measured value are below the reference yield stress; the lowest measured dimensionless yield stresses are 0.8 (f.i. set 4 at 30\% air content).

There is no major effect of Bio-Terge concentration nor bubble size on the dimensionless yield stress. During preparation however, we noted that incorporating aqueous foam in cement paste was more difficult for B2, where Bio-Terge concentration is lower, because the bubbles tend to break during hand mixing.

\begin{figure}[!ht]
\begin{center}
\includegraphics[width=5cm]{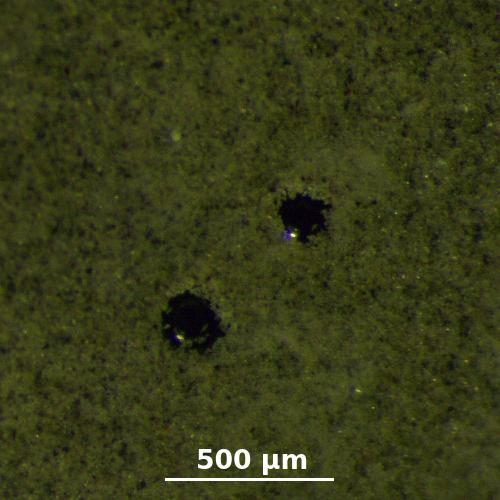}
\includegraphics[width=5cm]{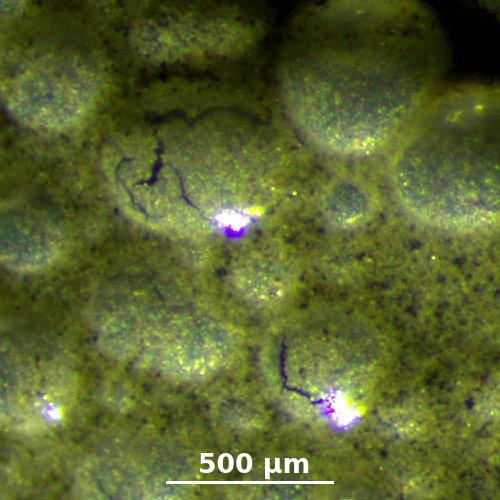}
\caption{Left: sample from set T1 containing TTAB. Right: sample from set B1 containing Bio-Terge. Note that bubbles tend to break rapidly when exposed to air, especially for samples containing TTAB, so that the number of the bubbles on each picture is not representative of the air content.}
\label{pictures_bubbles}
\end{center}
\end{figure}

Observation of the bubbles after they have risen at the top surface of the sample has revealed a fundamental difference for the surfactant samples: bubbles in Bio-Terge samples are covered with a layer of cement grains (Fig. \ref{pictures_bubbles}-right) whereas bubbles in TTAB are bare (Fig. \ref{pictures_bubbles}-left). Actually bare bubbles are difficult to observe because they usually break as soon as they reach the sample surface. On the contrary, in Bio-Terge samples at rest, only few bubbles break. The difference observed for the rheological behavior of the two surfactant samples seems to be related to the different nature of the bubbles surface. We consider each case in the following.

\subsubsection{Bare bubbles (TTAB)}
The effect of such bubbles are expected to be described by micromechanical models for capillary inclusions~\cite{2008_Chateau,2013_Nguyen}, which are based on the Bingham capillary number $Ca_y=\tau_{y,ref} R / (2 \gamma)$. Theoretical dimensionless yield stress for non-deformable bubbles $Ca_y\rightarrow0$ is given by~\cite{2008_Chateau,2013_Nguyen}:

\begin{equation}
\tau_y/\tau_{ref}=\sqrt{(1-\Phi)  \dfrac{5+3\Phi}{5-2\Phi}}	\text{ when } Ca_y\rightarrow 0
\label{equation_yieldstress_Ca0}
\end{equation}

When the bubbles are fully deformable ($Ca_y\rightarrow \infty$), the relation becomes:

\begin{equation}
\tau_y/\tau_{ref}=\sqrt{(1-\Phi)  \dfrac{3-3\Phi}{3+\Phi}}	\text{ when } Ca_y\rightarrow \infty
\label{equation_yieldstress_CaInf}
\end{equation}

\begin{figure}[!ht]
\begin{center}
\includegraphics[width=8cm]{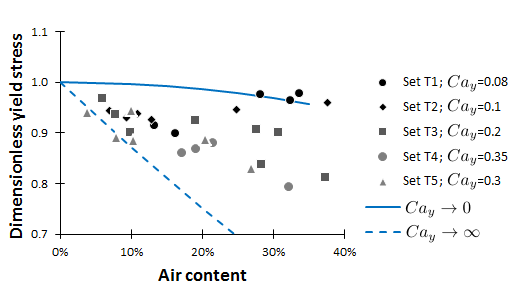}
\caption{Dimensionless yield stress for samples made with TTAB. Theoretical curves, which have been observed experimentally on model yield stress fluid by Ducloué~\cite{2014_Ducloue}, are indicated for non-deformable bubbles (full line) and fully deformable bubbles (dotted line).}
\label{graph_discussion_TTAB}
\end{center}
\end{figure}

Equations \ref{equation_yieldstress_Ca0} and \ref{equation_yieldstress_CaInf} are plotted in Fig.~\ref{graph_discussion_TTAB}. Both curves have also been obtained experimentally for bubble suspensions in a model yield stress fluid (concentrated oil-in-water emulsion) \cite{2013_Kogan,2015_Ducloue}. In \cite{2015_Ducloue}, experimental results fitted with equation \ref{equation_yieldstress_Ca0} for experimental Bingham capillary numbers within the range $0.0069 \leq Ca_y \leq 0.11$ and with equation \ref{equation_yieldstress_CaInf} for $Ca_y=0.57$.

Let us calculate $Ca_y$ for each TTAB data set. As mentioned previously, reference yield stress $\tau_{ref}$ decreases with increasing liquid content as described by equation \ref{equation_Yodel_c3}. Besides, at low surfactant concentration, surface tension $\gamma$ decreases when concentration increases. Such a variation is reported in~\cite{2017_Feneuil} as a function of TTAB concentration in the interstitial solution of the cement paste. 
By assuming that TTAB concentration in solution is close to the total TTAB concentration, i.e. that the amount of adsorbed molecules is negligible, surface tension can be described by $\gamma=31-6.8*ln(C_{res})$ within the range 0.02-0.2 g/L (this formula has been obtained by fitting experiemental results from \cite{2017_Feneuil}, with $\gamma$ in mN/m and $C_{res}$ in g/L). Using those $\gamma$ values and equation \ref{equation_Yodel_c3}, experimental values of $Ca_y$ are calculated and given in Table~\ref{table_Cay_TTAB}. In Fig.~\ref{graph_discussion_TTAB}, dimensionless yield stress results of aerated cement paste with TTAB are plotted for comparison with both theoretical curves.

\begin{table}[!ht]
\begin{center}
\begin{tabular}{|c|c|}
\hline
Set & $Ca_y$ \\ \hline
T1 & 0.08\\
T2 & 0.10\\
T3 & 0.21 - 0.23\\
T4 & 0.34 - 0.36\\
T5 & 0.28 - 0.30\\ \hline
\end{tabular}
\caption{Bingham capillary numbers for experiments with TTAB}
\label{table_Cay_TTAB}
\end{center}
\end{table}

At low air content,  experimental dimensionless yield stresses are systematically lower than theoretical values. This behavior is possibly an experimental artifact and shows that despite all our effort to keep the same preparation protocol for aerated and reference samples, shearing of the paste during hand mixing may be influenced by the presence of bubbles.

However, all the results fall within the range of expected values between the two theoretical curves. 
Moreover, the dimensionless yield stress values stretch as a function of $Ca_y$. This can be seen mainly at high air content: the higher the capillary number, the lower the dimensionless yield stress.

\subsubsection{Particle covered bubbles (Bio-Terge)}
Yield stress measured for aerated Bio-Terge cement pastes is well above the upper bound predicted by theory based on $Ca_y$ (i.e. equation \ref{equation_yieldstress_Ca0}). The particle layer attached to the bubble surface suggests that the behavior in the case of Bio-Terge lies in the modified surface properties of the bubbles.

Particle covered interfaces have first been reported by \textit{Ramsden}~\cite{1904_Ramsden} and \textit{Pickering}~\cite{1907_Pickering}. Bubbles covered by a dense layer of irreversibly adsorbed particles, called armored bubbles, are known to be very stable compared to surfactant stabilized bubbles: the particle armor can prevent bubble rupture (or coalescence)~\cite{1992_Denkov,2015_Timounay,2006_Kaptay} and volume decrease due gas transfer (ripening)~\cite{2006_Abkarian,2011_Stocco,2014_Maestro,2015_Pitois,2016_Taccoen}.

First requirement to allow solid particle adsorption at air-water or oil-water interface is that the particles must be partially hydrophobic. Then, the electrostatic barrier to particle adsorption must be low enough~\cite{2008_Tcholakova}. Increase of contact angle of water on cement grains has been observed in the presence of low amount of Bio-Terge, so we can assume that the first point is fulfilled. Cement grains have a size comprised between a few micrometers and 100~$\mu$m with average value close to 10~$\mu$m~\cite{2008_Mahaut_b}. Due to their irregular shape, radius of curvature of the edges is several orders of magnitude below the grain size~\cite{2010_Roussel}, which decreases the adsorption barrier~\cite{2008_Tcholakova}. In addition, cement solution contains a high concentration of dissolved ions~\cite{2015_Bessaies}, and this high electrolyte content reduces electrostatic repulsion between charged particles and interface, which also decreases the adsorption barrier~\cite{2008_Tcholakova}. All these aspects show that cement particles covered by a monolayer of Bio-Terge molecules are good candidates to adsorb at air-water interfaces.

In the micro-mechanical model for the bubbles, the stress of the air-fluid interface is characterized by one value, i.e. the surface tension $\gamma$. However, previous studies showed that this simple behavior cannot account for particle stabilized interfaces, where complex behavior arises from the kinetics and irreversibility of the adsorption of the particles at the interface \cite{2008_Tcholakova} and leads to an additional extra stress. In the case of an isotropic interface between two Newtonian fluids, the extra stress can be expressed as a function of to scalars, namely the shear surface viscosity $\mu_{s,s}$ and the dilatational surface viscosity $\mu_{s,d}$ \cite{1960_Scriven}. The resistance of surface viscosity to the shear stress from the bulk around the inclusion can be quantified by the Boussinesq numbers $Bq_s=\mu_{s,s}/(\mu_{ext} R)$ and $Bq_d=\mu_{s,d}/(\mu_{ext} R)$, where $\mu_{ext}$ is the viscosity of the external fluid. Determination of the value of the surface rheological properties of cement covered bubbles is far out of the scope of this paper. Though, the wide literature related to the rheology particle covered interfaces can highlight the behavior of adsorbed cement grains. We will recall here the main literature results. To get a more complete understanding of this complex issue, one can read the recents reviews from \textit{Thijssen and Vermant} \cite{2018_Thijssen}, \textit{Maestro} \cite{2019_Maestro} and \textit{Pitois and Rouyer} \cite{2019_Pitois}.

Adsorption of particles to and interface can enhance both the dilatational \cite{2017_Timounay} and the shear \cite{2014_Barman} surface viscosities. High surface viscosities indicates that under applied stress, deformation of the interface is slow. % These large Boussinesq values mean that interfacial mobility is strongly reduced with respect to bare interfaces
 When particle volume fraction is high, surface viscosities diverge and interface exhibits a yield stress behavior \cite{2016_Barman,2013_VanHooghten,2017_Beltramo,2014_Barman}, i.e. as long as the applied stress is below the yield stress, it behaves like a solid membrane. When interactions between particles are weak, yield stress behavior happens for particle surface fractions reaching the so-called \textit{jamming surface fraction} \cite{2017_Timounay}. On the contrary, when attractive interaction between the particles at the interface makes them agglomerate, the critical volume fraction leading to the yield stress behavior can be greatly reduced \cite{2016_Barman,2013_VanHooghten,2017_Beltramo,2014_Barman}. Note that for particle adsorbed at an interface, attractive interactions include capillary forces which are enhanced by the roughness of the particles \cite{2016_Barman,2013_VanHooghten}. Cement-grains covered interfaces are therefore likely to exhibit a surface yield stress.

In addition, the complex rheological properties affect the deformability of the bubbles. Recent numerical work performed on droplets in shear flow has shown that the deformation $D=(L-B)/(L+B)$ (where L and B are the major and minor axes of the deformed droplet) of the inclusion is greatly influenced by the surface viscosities~\cite{2016_Gounley}. For flow conditions which leads to a high deformation $D \geq 0.35$ when $Bq_s=Bq_d=0$, the deformation D is divided by a factor 2 when $Bq_s=Bq_d=10$ or when $Bq_d=0$ and $Bq_s=10$. For our system, the viscosities are expected to diverge, so we can assume that these values of the Boussinesq number are by far exceeded. Therefore, the bubbles must remain spherical during the start-of-flow experiment performed on the aerated cement pastes prepared with Bio-Terge.

Eventually, adsorption of hydrophobic cement grains at the surface of the bubbles leads to two effects: (1) the immobilization of the interface of the  bubbles due to a surface yield stress and (2) the reduction of the bubble deformation. Due to these two effects, armored bubbles are expected to behave as non-deformable solid inclusions.

\bigbreak

In order to compare directly the armored bubbles with solid inclusions, additional measurements were performed with solid beads instead of bubbles. Procedure followed for the tests is the same as described in paragraph \ref{part_protocol_c3}; at 27 minutes after preparation of the cement paste, we added to the paste 500~$\mu$m diameter polystyrene beads or 2~mm diameter glass beads. These additional samples contain no surfactant, so reference yield stress is given by equation~\ref{equation_Yodel_c3}.

The micromechanical analysis for solid inclusions assumes no-slip conditions at their surface~\cite{2008_Chateau}, and the corresponding theoretical dimensionless yield stress is:

\begin{equation}
\tau_y/\tau_{ref}=\sqrt{\dfrac{(1-\Phi)}{(1-\Phi/\Phi_m )^{2.5\Phi_m}}}
\label{equation_bead_suspension_micromechanical}
\end{equation}
where $\Phi_m$ is the maximal packing fraction of the beads within shearing conditions. This curve has been experimentally validated in model yield stress materials~\cite{2008_Mahaut_a} and in cement paste~\cite{2008_Mahaut_b}. $\Phi_m$ value measured in cement paste is 0.56.

\begin{figure}[!ht]
\begin{center}
\includegraphics[width=7cm]{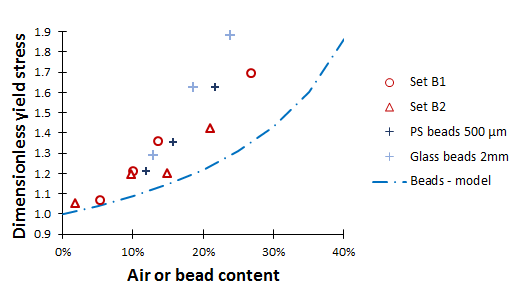}
\caption{Dimensionless yield stress for samples made with Bio-Terge. Dotted line correspond to the theoretical curve for a suspension of solid spheres with maximal solid fraction 0.56. Crosses correspond to measurements carried out with 500~µm polystyrene beads or 2~mm glass beads.}
\label{graph_discussion_BioTerge}
\end{center}
\end{figure}

In Fig.~\ref{graph_discussion_BioTerge}, we can see that our experimental results for Bio-Terge bubbles and for beads are consistent with equation \ref{equation_bead_suspension_micromechanical}, showing significant increase of the yield stress as a function of inclusion volume fraction, with no effect of inclusion size. Experimental data are slightly above the theoretical values, which could be related to our experimental method and the difficulty to follow accurately the same protocol for cement paste with and without inclusions. However, experimental dimensionless yield stresses for bead suspensions coincide with dimensionless yield stress of bubbles suspensions prepared with Bio-Terge surfactant. Therefore, armored bubbles behave mainly like solid spheres in the cement paste at the onset of yielding.

\section{Conclusion}

We have investigated the effect of added bubbles with well controlled size and volume fraction on the yield stress of aerated cement paste. Bubbles were stabilized with two surfactants, which are known to have different adsorption behavior with respect to cement grains. 

The behavior of the bubbles is strongly affected by the surfactant. When surfactant has low affinity to cement grains, the dimensionless yield stress depends only on the Bingham capillary number $Ca_y$, which accounts for the deformability of the bubbles. For $Ca_y \sim 0.1$, i.e. for small bubbles and/or low cement paste yield stress, the yield stress is almost unchanged with respect to the bubble-free paste. On the contrary, when $Ca_y \gtrsim 0.2$, i.e. for large bubbles and/or high cement paste yield stress, a small decrease of the dimensionless yield stress with air volume content can be observed.

Totally different behavior is observed when surfactant adsorbs on cement grains. Effect of bubbles is comparable to solid inclusions, and no effect of bubble size can be seen. We attribute this effect to the in-situ hydrophobization of the cement grains and their irreversible adsorption at the bubbles surface, which changes completely the surface properties of the bubbles. In such a case, the yield stress of the aerated cement paste increases significantly while its density decreases in proportion of air introduced in the paste.

\section*{Acknowledgements}

The authors want to thank Charles Joudon-Watteau for support to make the experiments.

This work has benefited from two French government Grants managed by the Agence Nationale de la Recherche [Grants number ANR-11-LABX-022-01 and ANR-13-RMNP-0003-01].

\section*{References}
\bibliographystyle{elsarticle-num} 
\bibliography{Bibliographie}

\end{document}